Designing Complexity? The Role of Self-Organization in Urban planning and Design

Authors: Anat Goldman & Efrat Blumenfeld-Lieberthal

The Azrieli School of Architecture, Tel Aviv University, Tel Aviv Israel

*Abstract*

This chapter explores the concept of self-organization in urban planning and design, highlighting its role in shaping the unique characteristics of cities. It examines how various socio-economic, cultural, and political factors contribute to the development of distinct architectural styles, emphasizing the morphological patterns and self-organization principles. The chapter addresses the emergence of scaling laws and fractal geometry in urban forms, using historical and contemporary examples to illustrate these concepts. The discussion also delves into the cognitive aspects of urban design, examining how the physical layout of cities influences cognitive maps and perceptions of urban environments, and how these perceptions, in turn, influence urban design. Through the prism of self-organization, it demonstrates the dynamic interplay between individual and collective actions and the shaping of the urban landscape. This analysis offers insights into the complex, self-organizing systems that define urban spaces, emphasizing the interdependencies among architectural design, urban planning, and human cognition in shaping cityscapes.

1. Introduction

This chapter introduces conceptually self-organization phenomena, i.e. the emergence of scaling laws in buildings and in open urban spaces. Cities are complex entities by their nature, and their landscape is affected by many socio-economic, cultural, political, and other influences and forces. They all contribute to processes of self-organization that result in the creation of unique architectural styles that characterize cities. These styles are often based on morphological patterns that govern the cityscape. This applies to the architectural design of buildings (i.e. the proportions of the built volumes, their positioning on the plots, the design of the façades, etc.) as well as to the design and the distribution of the open spaces in the city.

We start by introducing the term self-organization and explain in what way it is connected to architecture and urban design. Then, we demonstrate how self-organization is found in the morphology of individual buildings. Next, we present a cognitive model that explains how self-organization between the environment and human cognition creates a distinct city form. Finally, we provide real-world examples of this model in actual cities.

In 1961, Jane Jabocs published her canonical book "The death and life of great American cities" in which she criticized modern urban planning while presenting its effect on humans and urban life. In the last chapter of her book, she asks what kind of problem a city poses, and answers that cities are, in fact, problems of organized complexity where the numerous variables that create the city are not cluttered, they are "interrelated into an organic whole" (Jacobs, 1961; p. 565). In "A city is not a tree" Christopher Alexander (1965) compares cities to a semi-lattice, a topological structure of a set of elements and their connections. Indeed, as Bar-Yam (1998) points out in his book "Dynamics of Complex Systems" a dictionary definition of the term "Complex" is: "consisting of interconnected or interwoven parts" (p.1). This implies that in order to understand a complex system we must understand not only its parts but also their interrelations and interdependencies. Thus, complex systems can be understood as systems that contain numerous components and interactions between them, where the behavior of the system as a whole cannot be deduced from the behavior of its components (ibid). Complex systems can exhibit collective behavior that occurs due to the numerous interactions between the system's components. This collective behavior can be referred to as 'order' and the phenomenon where order appears due to the individual behavior of the system's components is called 'emergence'. This phenomenon is also related to one of the main attributes of complex systems, namely their self-organization. In the following, we explain the term 'self-organization' and demonstrate how it manifests itself in unique buildings as well as in cityscape, created by the assembling of numerous buildings.

## 2. Introduction to Self-Organization Phenomena

Self-organization refers to the spontaneous emergence of complex structures and patterns in a system without the need for external control or guidance. It is a process in which the components of a system spontaneously arrange themselves into a pattern or structure, often resulting in the emergence of new properties and functions at the global level. Self-organization is a common feature of many natural and artificial systems, and it can play a key role in the development and evolution of complex systems. Examples of self-organization can be found in a wide range of fields, including biology, physics, chemistry, computer science, and sociology (Karsenti, 2008; Perc 2013; Lehn, 2009; Mamei et al., 2006; Leydesdorff, 2001).

The term "self-organization" first appeared by Ashby (1947), though related ideas are considerably older. Krugman (1996) explained the multi-disciplinarity of the term. "What links the study of embryos and hurricanes, of magnetic materials and collections of neurons, is that they are all self-organizing systems: systems that, even when they start from an almost homogeneous or almost random state, spontaneously form large-scale patterns." (Krugman, 1996, p.3). He observed the way economies organize themselves in space and through time. Self-Organized criticality - SOC (Bak et al., 1988) describes a phenomenon where a system naturally develops into a critical state, where minor events can lead to a significant, often disproportionate response. SOC is characterized by several features including scale-free distribution of features, i.e. patterns within the system are similar at different scales. Their size distribution follows a power law equation indicating a scenario where small sizes are highly frequent while exceptionally large sizes are uncommon.

In 'Self-Organization and the City', Portugali (2000) describes self-organization as "the phenomena by which a system self-organizes its internal structure independent of external causes, is a fundamental property of open and complex systems. Such systems also exhibit phenomena of nonlinearity, instability, fractal structures and chaos – phenomena which are intimately related to the general sensation of life and urbanism at the end of the 20$^{th}$ century" (Portugali, 2000 p.49).

Self-organization in urban design is often expressed in terms of self-similarity. This refers to the repetition of resembling design elements or patterns at different scales within a city or a neighbourhood. In urban design, self-similarity creates a cohesive and coherent urban form, as the repetition of unique elements or patterns can help to create a sense of unity and order within the built environment. Notably, this phenomenon occurs despite the wide range of individual design choices made by various architects, urban planners, and residents. Each of these contributors, in their own unique and decentralized way, collectively shapes the city's design.

The above repetition can relate to building types and designs, street patterns, or public spaces. One representation of urban self-similarity is the fractal characteristics of certain urban attributes, where geometric shapes within the cityscape re-appear at different scales. In urban design, fractal patterns create complex and intricate patterns that sometimes resemble nature in terms of their aesthetics.

Self-similarity in urban design may create a sense of coherence and unity within a city or a neighbourhood, helping to create a sense of place. This can be especially important in large, complex urban environments, where the local repetition of certain elements can help to orient and guide people as they move through the city.

However, self-similarity may also lead to a lack of diversity in the built environment, as the repetition of certain elements or patterns can result in a monotonous and uniform appearance. This can be especially problematic in terms of the imageability of the city and wayfinding within it. To address this, urban planners usually use landmarks that create a sense of direction within the city.

One of the key principles of self-organization in urban planning is the idea that cities and neighbourhoods are complex systems that emerge and evolve through the interactions and decisions of individuals and groups. This means that the structure and form of a city or neighbourhood is only partly determined by a central authority or master plan, but to a large extent emerges spontaneously as a result of the decisions and actions of the people who live and work there. Thus, the collective behavior of the complex system can be found in the morphological and social characteristics of cities and the buildings within them. Here, we start by addressing the physical aspects of self-organization in buildings and cities that are expressed either by scaling laws or by the emergence of physical patterns that characterize a specific city.

Scaling laws are mathematical relationships that describe how a certain property or characteristic of a system changes as the size or scale of the system is varied. They are often expressed as power laws, in which the property of interest is proportional to some power of the size or scale of the system. Such scaling laws appear also in distributions. Unlike many other distributions, a power law distribution describes an uneven distribution, in which even extremely large values of the considered quantity can occur with non-vanishing probability. Power law distributions of different attributes of different systems describe self-similarities at different scales of the systems. For example, Zipf's law shows that the size of cities in different countries distributes according to a power law with an exponent close to 1. This means that the largest city is twice as big (in terms of its population) as the second largest city, three times the size of the third largest city, and so on (For more information about City size distributions see chapter 9 in this book).

Power laws, however, are not only limited to describing statistical attributes (e.g. population, income) but also physical geometrical forms such as fractals. A fractal, as defined by Benoit Mandelbrot (1982), is an object whose spatial form is not smooth in the sense of Euclidean geometry, thus termed "irregular", and whose irregularity repeats itself geometrically across scales.

The fractal geometry theory was conceived by the mathematician Benoit Mandelbrot in the 1970's. His famous study "The Fractal Geometry of Nature" (Mandelbrot, 1982) is considered the basis to this theory which complements Euclidean geometry. Mandelbrot coined the word fractal based on the Latin adjective "fractus." He chose this word because the corresponding Latin verb "frangere" which means "to break" or "to create irregular fragments." A fractal is a geometrical shape or pattern made of parts, which are similar to the overall pattern. It exhibits self-similarity, which means that each small portion, when magnified, can reproduce a larger portion (for elaboration see Chapter 8 in this book).

Batty and Longley (1994) demonstrated how fractal geometry might be applied to cities in many ways. Cities demonstrate fractal structure as their functions are self-similar across a variety of scales. When growing, cities do not occupy the area compactly. The gaps between built areas are the result of organic development as well as physical and planning restrictions.

While there is a consensus that cities are complex systems by their nature (Batty, 2007; Portugali, 2000), buildings, may not seem like natural complex systems, as they are composed of simple components (shapes, volumes, materials, and so on) that form a closed system. However, these shapes and volumes often do exhibit the characteristics of complex systems because their plans integrate numerous components including local regulations, building technologies and materials, different construction, design, and the personal preferences of the entrepreneurs and planners (architects and engineers that took part in the planning process).

And indeed, it was found that buildings also exhibit fractal characteristics of their plans, elevations, and volumes. It is important to note, that while mathematical fractals are infinite, similarly to fractals in nature, architectural fractals are not, as buildings are finite entities.

### 3. Fractal Geometry in planning and design

While self-similarity is an important concept, it is not always a feature that is created by the collective behaviour of people who live and use the city. In the next paragraphs, we address the use of fractal geometry intentionally or unintentionally by architects.

We can define fractal architecture as characterized by the repetition of geometric shapes at different scales, resulting in complex patterns. Such patterns are in analogy to those found in nature (e.g. trees, leaves, spirals). Many architects who incorporated such patterns into their architecture are said to be inspired by nature (see below). In addition to the aesthetic aspects of fractal architecture, there are also practical benefits that can be gained by incorporating fractal principles into buildings. The repetition of simple shapes at different scales usually creates strong and stable structures that are efficient in terms of material. This can be referred to as Biomimicry architecture or as Biomimicry models, followed by the application of their forms, processes, systems, and strategies in addressing human challenges.

Examples of fractal geometry in architecture can be found in the early monuments such as the Kandariya Mahadeva Temple in India (around 1000 AD) which is one of the Khajuraho temples that present self-similarity at different scales. Capo (2004) utilizes fractal analysis techniques to demonstrate that the classical orders possess a fractal-like property, characterized by the persistent presence of detailed patterns on increasingly finer scales (Joye, 2011).

Shapes that resemble well-known fractals are also found when examining medieval cities. Their surrounding walls often resemble the Koch Curve (Batty and Longley, 1994). Churches that were built during this time also present geometrical self-similarity in their façades, where the column, arches, and flying buttresses re-appear at different scales.

Moving forward to the Renaissance, we can find fractal characteristics in plans such as Michelangelo's and Bramante's designs for the Saint Peter's Cathedral, which also resemble the Koch curve (Hersey, 2001). Salingaros (2006) posits that ornamentation can play a crucial role in imparting self-similar characteristics to architecture by introducing detailed elements at even the smallest scales of the architectural form. And indeed, ornamentation is a key factor in the fractal-like characteristics of building styles like the Baroque and Rococo (1600-1800). During these periods, the façades of many churches in Europe presented fractal attributes where the main structure of the pillars that hold the pediments re-appeared at different scales (see figure 1). This structure can be found in the Jesuit church Il Gesù (Giacomo Della Porta, 1584), the Church of S. Susanna in Rome (1597–1603), the Church of St Peter and Paul in Krakow, Poland (1605–1619 to 1630) and many more.

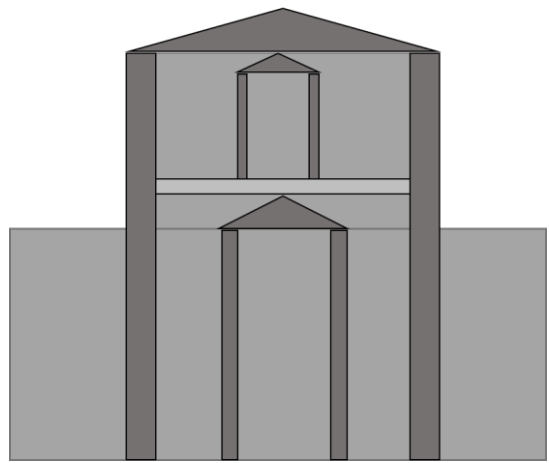

Figure 1: A general form of a façade of a Baroque church

Since fractal geometry is referred to as the geometry of nature, it is not surprising that buildings, designed by architects who were deeply affected by natural shapes and phenomena, are considered as having fractal characteristics. Frank Lloyd Write (1867-1959), one of the most famous architects, believed that natural relations comprise of "…the part is to the whole as the whole is to the part and which all devoted to a purpose consistently" (Meehan 1987). And indeed, his work is often referred to as containing fractal characteristics. Eaton (1998) posits that the fractal nature of the Wrght's Palmer House plan is apparent from the recurrence of an equilateral triangle as a geometric module at seven different scales, ranging from the complete structure of the building to specific elements within it. The major critique regarding Eaton's argument states that if the Palmer House plan is considered fractal, then any floor plan could be considered fractal as well, given that most houses contain spaces of various sizes which exhibit self-similarity as exemplified by the Palmer House plan (Joye, 2007). Harris (2007) agreed that the mere presence or repetition of similar shapes in a plan is not considered sufficient proof of fractal architecture, even in the work of an organic architect. He demonstrated the fractality of the Palmer House by developing a fractal shape that, by repeating it at different scales, after three iterations generates a similar pattern to the Palmer House plan (see figure 3 in Harris, 2007).

Approximately, at the same time, on the other side of the Atlantic Ocean in Barcelona, Spain. Antoni Gaudi started the Segrada Familia (1882). He was also known for being inspired by nature and nature's shapes. His masterpiece encompasses many different fractals. The arches are of different sizes, and the columns resemble tree trunks that divide into smaller and smaller branches. Staircases and ornaments such as spiral and flower-shaped fractals cover the interior of the church (see figure 2).

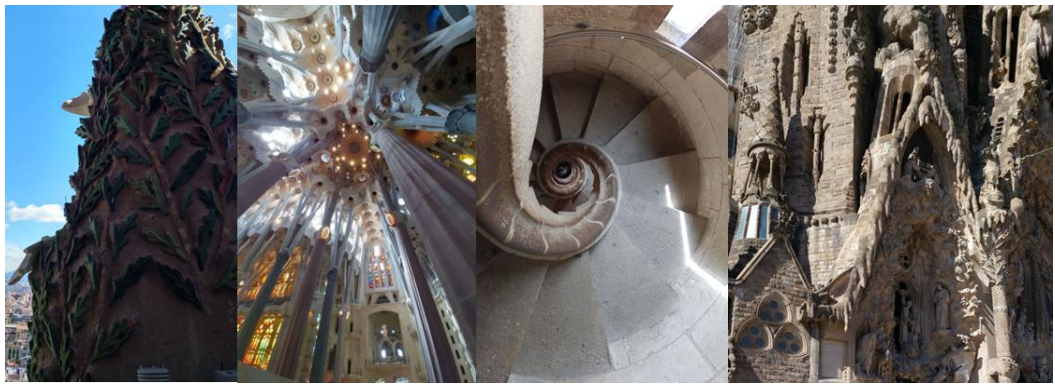

Figure 2: examples for fractals in the Sagrada Familia

In modern architecture, we find fractals as well. Despite the fact that modern technologies allow for new forms, we still find examples of using fractals in different types of architectural design. Both the *Tribunal judiciaire de Melun*, France (1992), designed by Jourda and Perraudin Architects, and the Stuttgart Airport Terminal, Germany (1991), designed by Von Gerkan, Marg+Partner, represent examples using a tree-like design for the constructional columns. A similar example can be found in the Tote Restaurant in Mumbai (2013), designed by the Serie Architects, UK.

Other examples, that are less explicit in the way they employ fractal geometry are the façade of the Grand Egyptian Museum, in Cairo (2014), designed by Heneghan-Peng Architects, where the façade is based directly on the pattern of the Sierpinski Pyramid; and the façade of the *Lideta Mercato*, Addis Ababa, Ethiopia (2017), designed by Xavier Vilalta which resembles the Sierpinski Cube. For further elaboration on fractals in architecture please see Rian&Sassone (2014) and Joye (2011).

4. Self-Similarity and Spatial Perception

As mentioned earlier, self-organization refers to the emergence of patterns in a system with no need for external or central control or guidance. After the first part of this chapter deals with the geometric expressions of self-organization in architecture, the second part will focus on the cognitive expressions of self-organization in urban design. While individual buildings usually present self-similarity in their morphology, cityscape is more complex, and its self-similarity is sometimes hidden from the eye, i.e. it is conceived only at the cognitive level, because the scale of an urban area is too large for a human eye to capture at once.

The urban street network serves as an example. When visiting a new city, we usually study the main streets and the network they form to help us navigate. Although these streets constitute only a small fraction of the total streets in a city, they are sufficient to help us create a mental map that helps us find our way. Jiang (2007) claimed that less than 1% of the streets in a city, which are the most connected ones, are sufficient to create the city's skeleton. In later works (Jiang 2013a, 2013b), he claimed that the image of the city (which can be referred to as a mental map) can be generated from the perspective of the different elements that form the city. He suggested that the image of the city is the outcome of the scaling laws that govern size distributions of urban artifacts or locations. But what are these mental maps that Jiang refers to? And how do they emerge?

Tolman (1948) introduced the concept of "cognitive map," which refers to the internal representation in the mind of a living organism (human or animal) of their external environment. These maps aid the organisms in navigating and exploring their surroundings. Lynch's "The Image of the City" (1960) showed how people perceive urban environments. He proposed that the image of a city is composed of five elements: edges, nodes, paths, districts, and landmarks. He also claimed that the image of the city is based on three attributes: identity, structure, and meaning. Identity differentiates the city from other entities, structure refers to the spatial or pattern relationship with the observer/other objects, and meaning can be either practical or emotional to the observer.

Portugali (1999, 2011) defines the relationship between people and the environment as an inter-representational network (IRN), composed of external and internal, cognitive and emotional representations. These representations are linked in a feedback loop, so that external representations become internal through sensory absorption and internal representations become external through verbal, visual, and gestural expression. To explain the events within the cognitive system itself, Portugali draws on the theory of synergetics represented by the letter "S" in a new model which he calls SIRN (Haken, 1987). According to this theory, awareness, cognition, cognitive mapping, and interaction between internal and external representations are all complex systems that develop in line with synergetic principles in self-organization.

Cognitive map research shows that each person has a unique representation of the same space. This representation is related, among other things, to early familiarity with the place, modes of arrival, early social perceptions, and so on. The cognitive process involves interactions between internal representations that continue to form and change over time, according to the level of familiarity with the environment, and to produce a new external representation (Portugali, 2000). Thus, the internal and external representations are described as ad-hoc identities created from the dynamics of the system and the context.

The cognitive process that leads to our spatial perception is dynamic and leads to the expression of multiple representations according to the task at hand. The process is described by Haken and Portugali (1996) in a variable-based general model that portrays the effect of information acquisition by an agent, its processing, and the dissemination of the processed information.

The model is described as consisting of three sub-parts that relate to the number of agents involved in the process. The first refers to an individual, the second to a meeting of a number of agents, and the third to the interaction between a large number of agents connected to a collective information repository (also referred to as culture or society).

Regarding the current chapter, a third model, the Interpersonal-Information Reservoir Model, is of particular significance. This model describes a cognitive-collective process involving a large number of agents. The importance of this process is due to an understanding that every agent's action in the urban environment contributes to redesigning the information that the city provides to other agents, i.e. commuting to work, moving to a new house, shopping, constructing new buildings, building infrastructure, etc. In this sense, the city can be viewed as a common reservoir produced collectively by the actions of many individuals, including collective architecture. This reservoir has a physical expression in a variety of forms, i.e. text, objects, buildings, and even whole cities.

In his book "Invisible Cities" (1978), Italo Calvino outlines the description of cities in the conversation between Marco Polo and Kublai Khan. Polo portrays the city Zoe as follows:

"The man who is traveling and does not yet know the city awaiting him along his route wonders what the palace will be like, the barracks, the mill, the theatre, the bazaar. In every city of the empire every building is different and set in a different order: but as soon as the stranger arrives at the unknown city and his eye penetrates the pinecone of pagodas and garrets and haymows, following the scrawl of canals, gardens, rubbish heaps, he immediately distinguishes which are the princes' palaces, the high priests' temples, the tavern, the prison, the slum. This, some say, confirms the hypothesis that each man bears in his mind a city made only of differences, a city without figures and without form, and the individual cities fill it up." (ibid., p. 34).

Calvino's perception of the image of the city resembles Portugali's model as he too, describes an inner representation of a general notion of the city we carry within us.

Blumenfeld-Lieberthal et al. (2018) tried to extract the common image of a city that we all bear in our minds. They emphasized the city's morphological structure, with a focus on its statistical characteristics and how they are perceived. They ran a survey where they asked the participants to rank different configurations of artificial urban morphologies in order to identify the morphological attributes that form the inner representation of cities.

The artificial environments they used were created using a computational method where they controlled and changed different morphological characteristics. These environments were based on two basic street-networks: a non-orthogonal network that corresponds to European cities that were developed bottom-up (i.e. without an overall urban plan), and an orthogonal grid, which corresponds to North American cities that were developed top-down (see Alexander, 1965 for elaboration). Additionally, they used two different size distributions that obeyed either power-law distributions (with different exponents) or Poisson distributions (with different values of λ). Environments created with power-law distributions exhibited stark differences between largest and smallest sizes (e.g. street length, building height), when the exponent was large, and moderate difference when the exponent was small. Power law distributions, when referring to the morphology of the urban entities, are related to fractal morphology. By doing so, the authors examined weather environments that resemble organic morphologies are perceived as resembling urban environments (which are actually man-made artifacts).

The Poisson distribution created a rather uniform set of sizes, where the majority of entities are near the average size, and there is a sharp decline in sizes at both the smallest and largest ends, leading to a limited range of sizes. In the present context, a high value of λ leads to a more uniform distribution of sizes, characterized by a narrow size range, whereas a low value of λ expands this range and results in higher values for entity sizes.

The authors found that orthogonal street grids were the most significant predictor of an environment's similarity to an urban setting. However, power-law distributions of entity sizes (of street length and width as well as building height) can also serve as predictors. The combination of these predictors of urban similarity suggests that the mental image of a city is not defined by a single feature, but rather by a combination of top-down planning (represented by the orthogonal grid) and bottom-up processes (represented by the power-law distributions). These findings extend Calvino's (1978) and Portugali's (1996) 'inner representation of the city', as it seems that not only the differences between the urban entities form our mental image of the city, but also the morphological and statistical relations between them.

Imageability, a concept introduced by Kevin Lynch in his influential book "The Image of the City" (1960), refers to the quality of a physical environment that makes it clear, memorable, coherent, and recognizable to people. Lynch argued that the imageability of the urban environment is essential in helping individuals forming a mental map of the city, enabling them to navigate and understand the urban space successfully. Shushan et al. (2016), used virtual reality to test which morphological configuration creates better imageability of urban areas. They too showed that heterogeneous environments (e.g., environments with fractal characteristics) offer higher levels of information and greater imageability compared to homogeneous environments, which are characterized by normal distributions.

To summarize the significance of the SIRN model, we can first claim that it identifies that the human cognitive system can be conceptualized as a complex and open system with characteristics of emergence and self-organization. In other words, complex systems operating in self-organization are present at various levels, ranging from individual human cognition to the relationships between a group of people in a small community or in a society, and the urban environment, which is also an open and complex self-organizing system. Second, it emphasizes the dynamic process in which urban landscapes are actively created and changed over time, due to the complexity of the process, involving the generation of individual products, connections between individuals, and ultimately the collective action of society in a constructive process.

## 5. Self-Organization and Cityscapes

Self-organization is also present in cities given the way we perceive them as many cities have a unique style that characterizing them. When we see the cityscape of cities like New York, Paris, or Jerusalem we can usually identify these cities even if we haven't visited them. The question – of what makes a place look like itself – occupied researchers from various disciplines. Doersch et al. (2012) approached this question from a computer science point of view and developed a method to automatically identify the most distinctive elements that form the image of a specific area (in their example – the city of Paris). They claimed that the "look and feel" of a city largely lies in a set of stylistic elements that are consisted of the visible details that we are exposed to in our daily urban life, rather than on the few famous landmarks (that correspond to Lynch's elements). While we may assume the Hausmann Boulevards, The Eifel Tower, or The Louvre might be the best representative of the city, the authors found that it is rather the balconies, windows with their railings, and the traditional blue/green/white street signs which were the most informative features identifying the city as Paris. This can be explained as a self-organized phenomenon related to the SRIN model presented earlier.

Portugali (2011) approached this issue from another perspective and explained the exemplary phenomenon of the butterfly effect of balconies in Tel Aviv as self-organization. The city of Tel Aviv, established in the early 1920s, in characterized by many Bauhaus buildings with verandas. Once the first resident decided to expand their apartment by closing the balcony and converting it into a half-room, others followed him/her, and the cityscape of Tel Aviv (and later other cities in Israel) changed. To counteract this trend, the municipality of Tel Aviv imposed taxes on all balconies, regardless of whether they were open or closed (unlike previously when only closed balconies were taxed). This caused developers to construct buildings with closed balconies, as they enlarged the potential size of the apartments. However, with the advent of postmodern architecture, balconies regained popularity as decorative features. To meet the demands of this changing architectural style, city planners granted permits for the construction of open balconies, which could not be closed, leading to the prevalence of "jumping balconies" in the urban landscape of Israel (Figure 3). This style gained popularity to the stage that a new hotel in Tel Aviv was built with jumping balconies (even though it is obvious that none of their guests will ever close the veranda in order to enlarge their rooms).

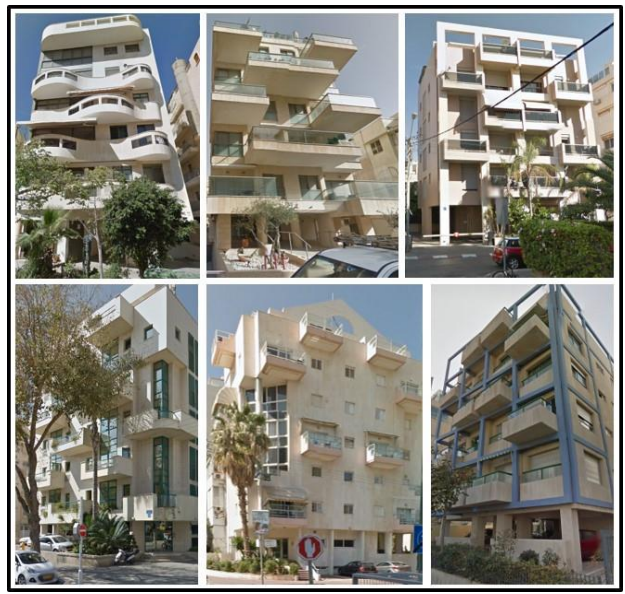

Figure 3: Examples for jumping balconies (Goldman, 2022)

Goldman (2022) followed Haken and Portugali's SIRN model (1996), in order to identify common characteristics of shared residential buildings in Tel Aviv. These types of buildings constitute the majority of buildings in central Tel Aviv and therefore represent reliably the common and mundane architecture of everyday life. Based on the assumption that architects, as well as residents, are affected by the environment they plan within, a local appearance of the city can be detected.

Goldman (2022) found that the typical building has several repeating characteristics. First, it has a box shape of three to four stories; second, is built on pillar-supported level; and third, balconies are a routine feature appearing on every floor on the main façade. Thus, creating a vertical sequence. She grouped the residential buildings into several classes and sub-classes, based on the topology of their façades. The different typologies can be examined by grouping them into classes with similar characteristics. For example, Group A includes type A and a, Group B includes types B and b, etc. (See

Figure 4). Goldman found that the balcony represents the recognizable symbol of the Tel Aviv residential building (as 100 percent of the buildings she studies had balconies). Although the balcony characterizes the building in a Mediterranean environment it is not a local invention (Fox, 1998). It has become as a key component in Tel Aviv since its establishment, and balconies facing the main street characterize each of the different types of residential buildings. In addition to the climate role of the balcony in ventilating the apartment, it also serves as a key design element in creating the building's primary façade. The reason for this is that the buildings, being densely positioned in the city center, create a hierarchy in the architectural treatment of the façades. Usually, only one façade of the building faces the street (with the exception of corner buildings which have two street-facing sides). The sides of the buildings only offer a view on the front façades and most are obscured by the plot's depth, with the rear façade not visible from the street. Hence, the main design focus of architects is on the street façade, and balconies play a key role in creating visual depth and serving as a focal point. Despite the city's efforts to direct the design of balconies and prevent their enclosure, the appearance of balconies in the 1930s was diverse. In practice, balconies were built in various configurations without considering legal restrictions but rather based on the accumulative everyday actions of residents that preferred converting the balcony's open space into additional in-house space.

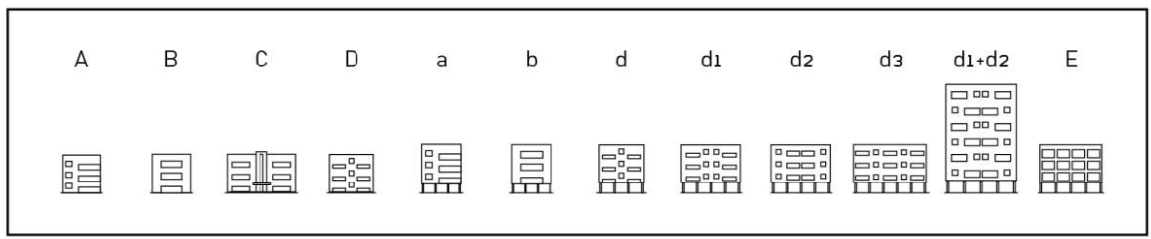

Figure 4: The typologies of buildings characterizing the center of Tel Aviv (Goldman, 2022)

In addition to the balconies being a key element in shaping the cityscape of Tel Aviv, the Pilotis were another key element that appeared in most of the façades of Tel Aviv from 1939 onwards. Figure 4 shows that 75% of the different typologies that Goldman (2022) found in the Tel Aviv residential buildings have pilotis that elevate the buildings from the ground floor.

The use of columns (pilotis) in construction arose mainly due to the technological revolution in the building industry that took place at the end of the 19th century in Europe and the US, which involved the adoption of reinforced concrete as the primary building material. In Tel Aviv, the adoption of reinforced concrete started in the early 1930s, with 1933 being a significant year. Building on columns was a ground-breaking and aesthetically pleasing concept (following Le Corbusier's pilotis in villa Savoye) that allowed for the separation of the building mass from the ground and the increase in height of heavy structures. This resulted in the creation of an "air pocket" between the ground level and the street and within the building mass itself, which now starts 2.5 meters above the street level on the first floor. Additionally, using reinforced concrete in columns, beams, and girders offered advantages for subsequent floors of the building, such as reduced wall thickness, leading to larger spaces, and wider and larger openings (especially windows) compared to before.

However, at first, the idea of building with columns was not favoured by the municipality and it took several years until it was legally established in early 1939 after the number of requests to integrate columns into the ground floor of residential buildings had increased. The decision, adopted by the municipality to embed columns at the ground level facing the street was a compromise that blended modernist concepts, as represented by architects, with the city's goal of reducing the necessity of enforcing building code violations. This approach was adopted in recognition of the continued occurrence of these violations, and the belief that it was better to address them in a straightforward manner. Resembling the balcony affair, here too the municipality had to adopt new regulations to address the order that emerged bottom-up (i.e. self-organized), from planners and people who lived in the city. Despite the evolution of building materials, urban regulations, and architectural styles, elevated buildings on pilotis are also built in Tel Aviv today, and still represent the cityscape.

## 6. Summary

To summarize, cityscape that forms the image of cities in our minds is created by self-organizing processes that combine bottom-up forces (e.g., architects, planners, and residents of cities) with top-down ones (e.g. urban regulations). The results of these forces and the processes they lead to are the unique styles of different cities, that are expressed at different scales, i.e. the building design, the size distribution of streets length and width, etc.

In this chapter, we reviewed self-similarity and self-organization in both buildings and cityscape and showed that while the representations of these phenomena are usually morphological, they affect our spatial perception of our surroundings and create a circular causality where the built environment affects the way we act within it, and the way we act within it affects, in return, the built environment. This complexity is typical for urban systems at different scales, and their relations with humans.

### Acknowledgement

The authors thank Prof. Juval Portugali for insightful conversations and co-tutoring Anat's PhD research.